\documentclass[a4paper,10pt,twoside]{cpc-hepnp}
\usepackage{CJKutf8}
\usepackage{upgreek,fancyhdr}
\usepackage{multicol}
\usepackage{graphicx}
\usepackage{booktabs}
\usepackage{amssymb,bm,mathrsfs,bbm,amscd}
\usepackage[tbtags]{amsmath}
\usepackage{lastpage}

\usepackage{siunitx}

\usepackage{comment}

\begin{document}
\begin{CJK*}{UTF8}{gbsn}

\fancyhead[c]{\small Chinese Physics C~~~Vol. xx, No. x (201x) xxxxxx}
\fancyfoot[C]{\small 010201-\thepage}

\footnotetext[0]{Received 31 June 2015}

\title{A versatile PMT test bench and its application in the DAMPE-PSD\thanks{Supported by the Strategic Priority Research Program of the Chinese Academy of Science under Grant No. XDA04040202-3}}

\author{%
      Yong Zhou (周勇)$^{1,2,3}$
\quad Zhi-Yu Sun (孙志宇)$^{1}$
\quad Yu-Hong Yu (余玉洪)$^{1;1)}$\email{yuyuhong@impcas.ac.cn}
\quad Yong-Jie Zhang(张永杰) $^{1}$\\
\quad Fang Fang (方芳)$^{1}$
\quad Jun-Ling Chen (陈俊岭)$^{1}$
\quad Bi-Tao Hu (胡碧涛)$^{2}$
}
\maketitle

\address{%
$^1$ Institute of Modern Physics, Chinese Academy of Sciences,  509 Nanchang Road,  Lanzhou,  730000,  P.R.China\\
$^2$ School of Nuclear Science and Technology,  Lanzhou University,  222 South Tianshui Road,  Lanzhou,  730000,  P.R.China\\
$^3$ Graduate University of the Chinese Academy of Sciences,  19A Yuquan Road,  Beijing,  100049,  P.R.China\\
}

\begin{abstract}
A versatile test bench system, dedicated for massive PMT characterization, is developed at the Institute of Modern Physics, Chinese Academy of Sciences. 
It can perform many test contents with large capacity and high level of automation, and the migration from one testing configuration to another is lightweight and time-saving. 
This system has been used in the construction of the Plastic Scintillator Detector of DArk Matter Particle Explorer already, and a total of 570 Hamamatsu R4443 tubes have been tested successfully.
\end{abstract}



\footnotetext[0]{\hspace*{-3mm}\raisebox{0.3ex}{$\scriptstyle\copyright$}2015
Chinese Physical Society and the Institute of High Energy Physics
of the Chinese Academy of Sciences and the Institute
of Modern Physics of the Chinese Academy of Sciences and IOP Publishing Ltd}%

\begin{multicols}{2}

\section{Introduction}
\label{sec:introduction}

Since invented about 80 years ago, Photomultiplier tubes (PMTs) have been widely used as photosensors in nuclear and particle physics experiments due to the high sensitivity, fast time response and other benefits they have. 
Today, large experiments may contain thousands or even more PMTs. To make the detectors work well and get optimized performance, it's important to study the characteristics of each PMT before they are put into usage. 
Manufacturers provide datasheets with typical values of certain parameters that are normally used in the industry, but may not be useful or applicable to a particular science experiment
On the other hand, the experiments may have special requirements for some characteristics of the PMTs, which the manufacturers may not provide. 
Thus characterization of PMTs in the laboratory is mandatory in large physics experiments.

Testing a large number of PMTs in limited time is a tough job, and usually a dedicated test bench is constructed to facilitate this work~\citep{barnhill_testing_2008,akgun_complete_2005,adragna_pmt-block_2006}.
Setting up a system like this is not a trivial work, which demands considerable investment of time and effort.
On the other hand, testing of PMTs is a commonly encountered procedure and many components of the testing configuration can be shared among different applications.

In this paper, a versatile test bench dedicated for PMT characterization is reported.
The test bench is designed to be a standard laboratory equipment for fast PMT characterization for several experiments prepared and planned at the Institute of Modern Physics (IMP), Chinese Academy of Sciences (CAS).
To accommodate the various requirements of different experiments, the test bench adopts a modular design pattern both in the hardware and the associated software platform.
This makes the migration from one testing configuration to another light-weight and time-saving.

The first user of this test bench is the Plastic Scintillator Detector (PSD), a sub-detector of the DArk Matter Particle Explorer (DAMPE)~\citep{Chang_Jin_dampe} which is a satellite-borne experiment. 
The test work for the PMTs of PSD using this test bench is very successful. Summary of the test results are shown in Sec.~3.

\section{Description of the test bench}
\label{sec:description}
	
A schematic diagram of the PMT test bench is shown in Fig.~\ref{fig:FIG1}, it is designed to test up to 25 PMTs simultaneously.
Light pulses from a blue LED are distributed to the PMTs through an integrating sphere and a transparent fiber bundle, which are mounted on a three-axes motorized stage.
PMTs under test will be mounted on a separate fixed stage, thus allowing for photocathode's position scanning of all tubes simultaneously.
There are also two PMTs fixed on the motorized stage, serving as a reference to monitor the stability of the LED as well as the performance of the whole system.
Both stages are housed in a light-tight container made of aluminum alloy, with a dimension of $176cm\times100cm\times78cm$.

\end{multicols}
\ruleup
\begin{center}
	\includegraphics[width=140mm]{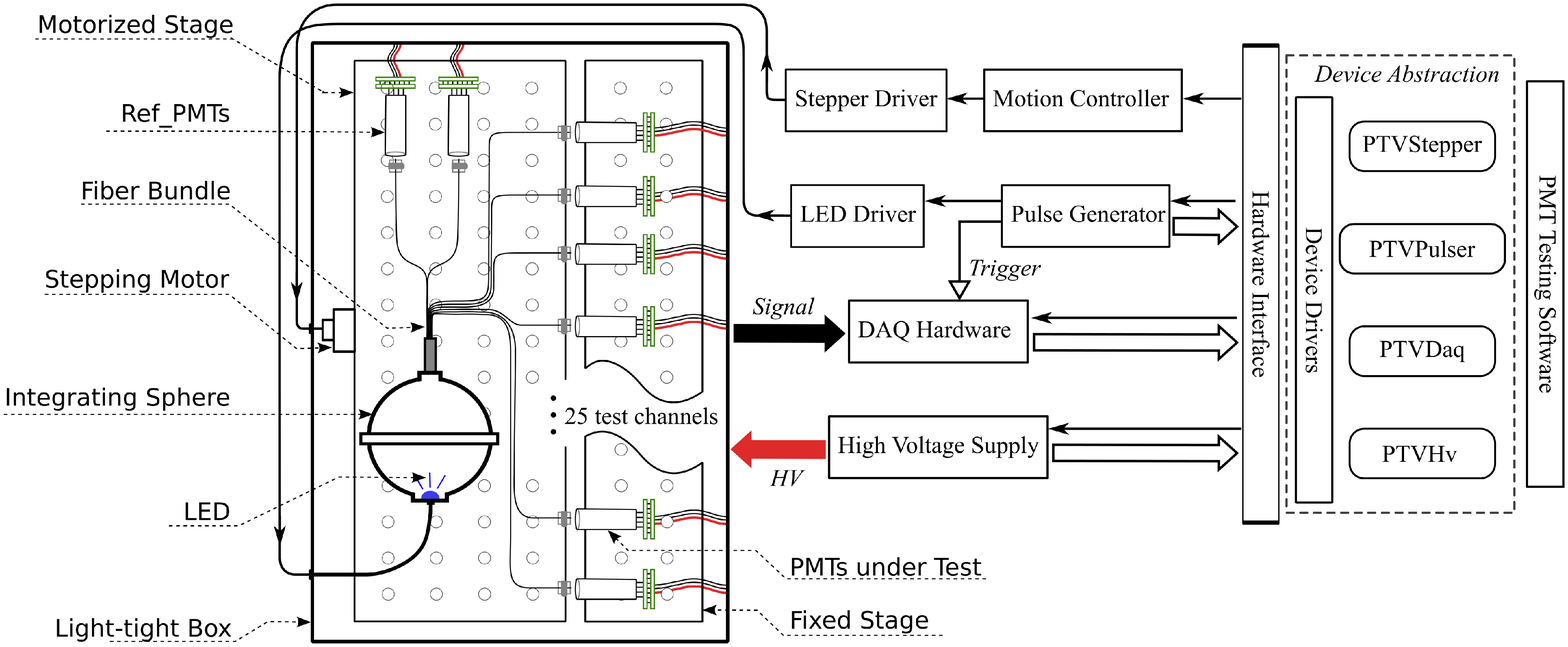}
	\figcaption{\label{fig:FIG1}Schematic diagram of the PMT test bench system.}
\end{center}
\ruledown
\begin{multicols}{2}
	
For the proper operation of the test bench, a variety of auxiliary equipments are needed. 
They are divided into four groups according to the functions they provide: motion controller, pulse generator, high voltage supply and data acquisition system(DAQ).
The motion controller and the pulse generator are tightly coupled to the test bench itself, thus the same hardware can be reused for nearly all the cases the bench may work for.
On the other hand, the DAQ and the high voltage supply system have closer relation to the underlying experiment which need massive PMT characterization, and normally project-specific hardware is preferred in different experiments. 
Originally, a universal CAEN SY1527LC~\citep{sy1527lc} power crate is adopted as the high voltage supply system and a CAMAC crate with CC-USB crate controller~\citep{cc_usb} as the DAQ.

Auxiliary equipments are placed outside the light-tight container, and the cables are led out through light-tight feedthroughs.
The control of all these equipments are integrated into a single software. The changing of hardware is handled smoothly in the software design and automation is realized in every aspect of the test bench.

The whole test bench is sitting in a cleanroom at IMP, and the room temperature is kept to be $22\pm2$\si{\celsius}~all the time. 
In the following sections, some key components of the test bench will be described in detail.

\subsection{Motorized and fixed stages}
\label{sec:stages}

The motorized and fixed stages are the main body of the test bench.
All other objects inside the light-tight container are mounted on the top of them.
In particular, customized fixtures for fibers, PMTs and integration sphere have been designed for convenient and accurate positioning.

\begin{center}
	\includegraphics[width=\linewidth]{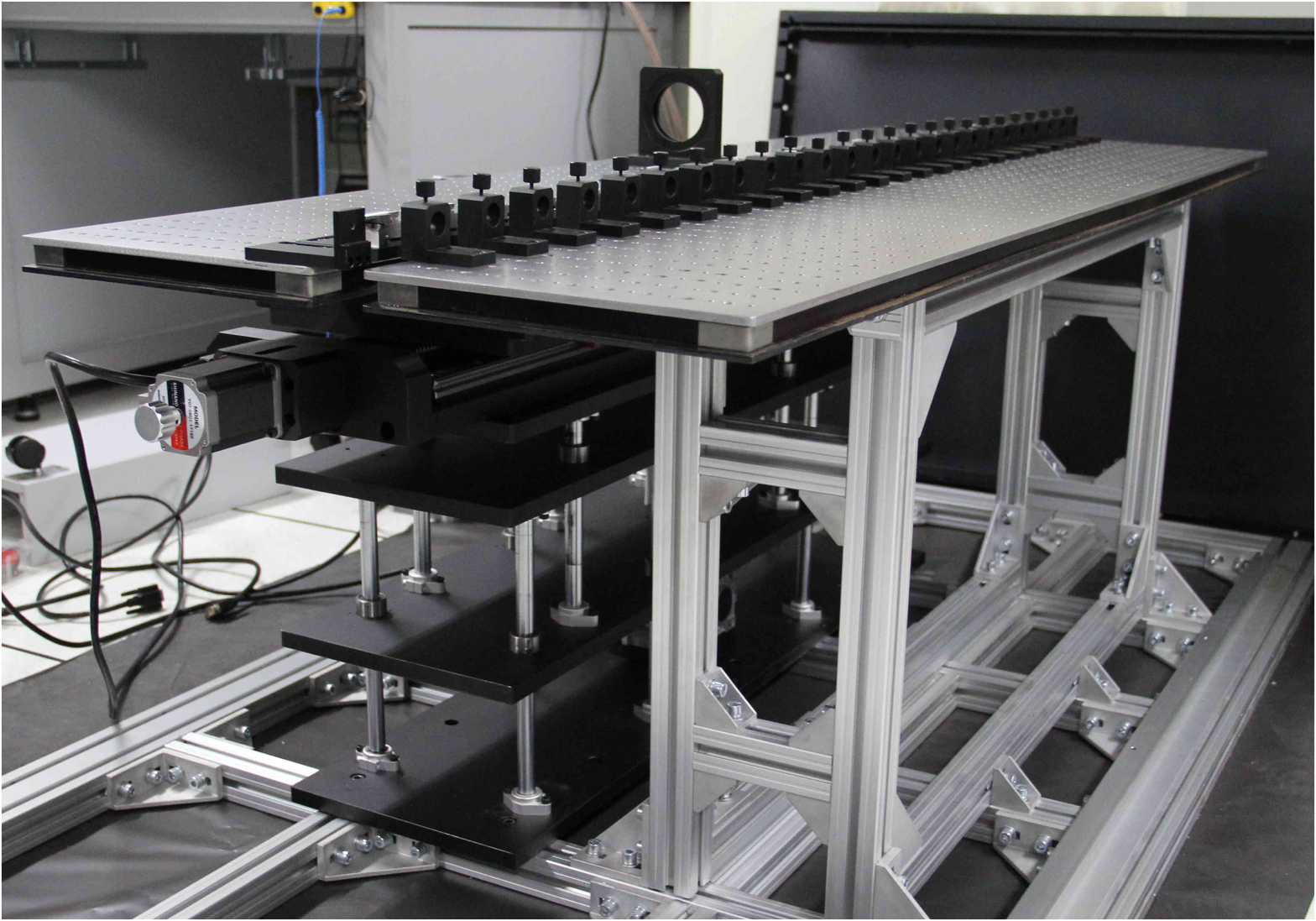}
	\figcaption{\label{fig:FIG2}Motorized and fixed stages before assembly, with the fixtures for the integrating sphere, fibers and PMTs on the top.}
\end{center} 

As shown in Fig.~\ref{fig:FIG2}, both stages are covered with a $1560mm\times250mm$ optical breadboard. 
These breadboards are made of \SI{2.5}{cm} thick stainless steel, providing substantial resistance to deformation in this application. 
The grid pattern of tapped holes on their surface provide extra flexibility in the testing configuration as well as facilitates mounting/unmounting operations.

The load capacity of the motorized stage is \SI{30}{\kilo\gram} and it can do a three-dimensional motion with the help of three stepping motors, which have a minimum step of \SI{1.56}{\micro\meter}.
The stage can move up to \SI{60}{\milli\meter} horizontally and \SI{70}{\milli\meter} vertically, and it's enough to cover any PMTs smaller than 2 inches which are mostly used.
The stage can also move along the third direction for $15mm$, this is to protect the fibers while mounting/unmounting the tubes as well as to control the gap between the fiber and the input window of PMT.

\subsection{Light source}
\label{sec:light_source}

A high-power blue LED~\citep{z-light}(\SI{3}{\watt},  \SIrange{465}{485}{\nano\meter}) is adopted as the light source of the test bench.
The LED is fixed on a special designed base using thermally conductive silicone rubber, and then coupled to a  \SI{5}{\centi\meter} diameter integrating sphere~\citep{integrating_sphere} directly.
A uniform light source can be achieved in this way, and the uniformity could make the coupling between the light source and the fiber bundle much more flexible.

For PMT characterization, light pulses of short duration are needed.
The general purpose pulse generator, Tektronix AFG3252~\citep{afg3252}, is adopted to drive the LED directly.
AFG3252 can adjust all the pulse parameters in a wide range with high precision. This is a critical feature for a test bench with an objective of versatility, as diverse requirements for the light pulses exist in different applications. 
Besides the LED driving pulse, AFG3252 can also output a synchronized pulse as the trigger signal to the DAQ, which  simplifies the DAQ configuration in most cases. 

The uniformity of the light source has been checked by using the same optical fiber to scan the output port surface of the integrating sphere. The fiber was coupled to a PMT to measure the light intensity, and an uniformity within $\pm\SI{0.5}{\percent}$ has been reached.

\subsection{Fiber bundle}
\label{sec:fiber_bundle}

A bundle of 35 plastic clad silica fibers~\citep{optical_fibre}(25 for test channels, 2 for reference channels and 8 spares) is utilized to distribute light pulses to each PMT.
Each fiber is \SI{1.5}{\meter} long and has a \SI{400}{\micro\meter} diameter core with a \SI{75}{\micro\meter} thick cladding, and the numerical aperture(NA) is 0.37.
The relatively large core and NA make the fiber an efficient light extractor of integrating sphere output. 

To protect the fibers from mechanical damage, both ends of the fiber bundle are coated with stainless steel ferrules.
The bundle end is coupled to the center of the output port of the integrating sphere using a fiber alignment stage.
On the other end, each fiber is fixed using a customized fiber holder which allows two-dimensional position adjustment, and is aligned to the center of the corresponding PMT input window with a precision of \SI{0.5}{\milli\meter}.

The light output difference between fibers has been calibrated after coupling the bundle to the integrating sphere.
The same PMT and light pulser setting was used to measure the light output of each fiber successively.
The PMT was fixed and each fiber was aligned to the same position with a precision of \SI{0.5}{\milli\meter} using a fiber alignment stage, and the light intensity fluctuation of the light source was monitored by a separate PMT.
After the light intensity correction, a variance of \SI{10}{\percent} is observed among the fibers.
Different light intensities were used in the calibration process, and the result shows that the light output difference has no dependency on the light intensity as shown in Fig.~\ref{fig:FIG3}. 
Thus, the light output difference can be treated as a constant parameter and the results will be utilized in the measurement of the relative gain of PMT (see Sec.~3.1)

\begin{center}
	\includegraphics[width=\linewidth]{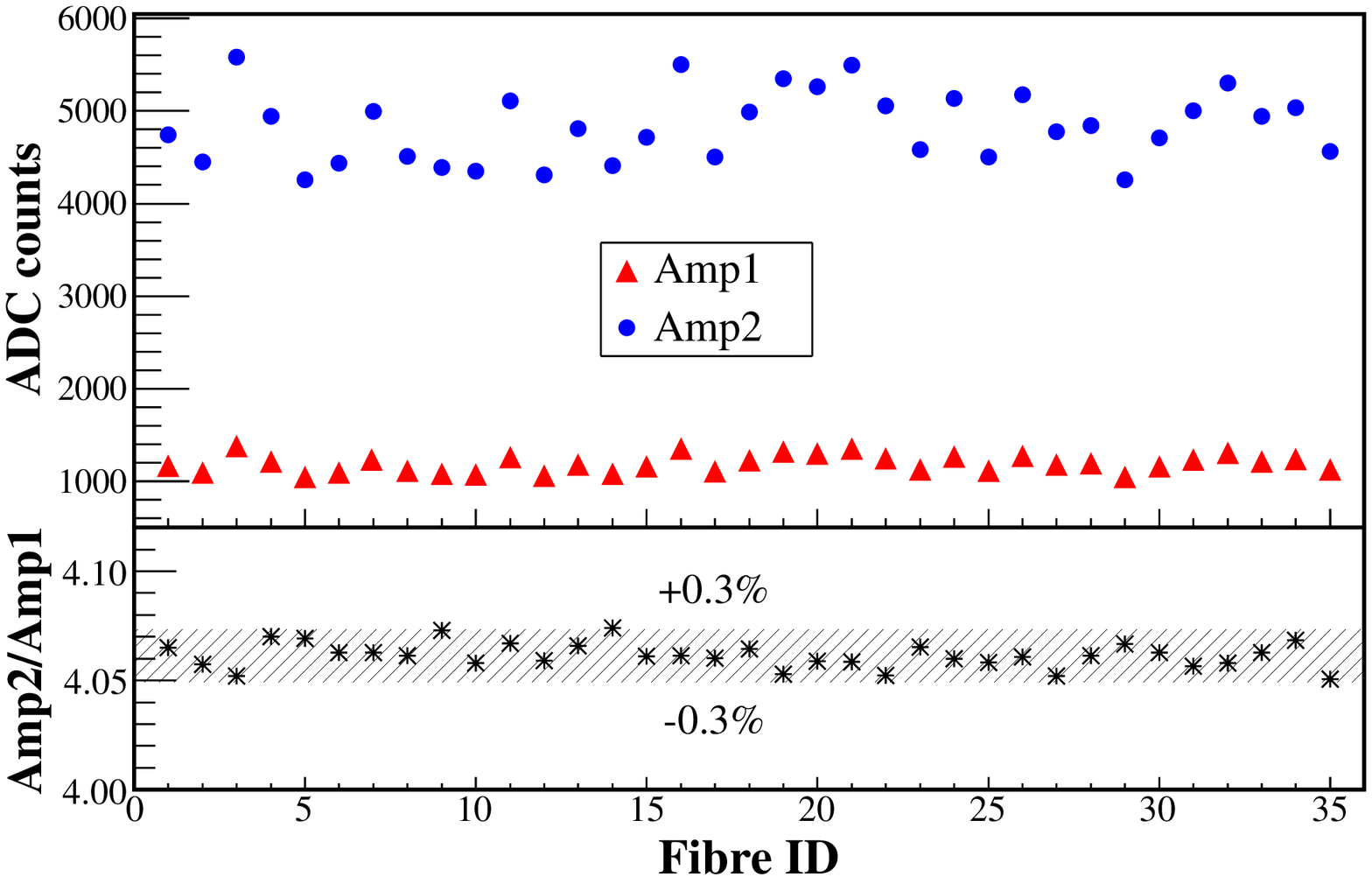}
	\figcaption{\label{fig:FIG3}Up: Light output of 35 fibers measured at two light intensities with an amplitude difference of about 4 times.
		Bottom: Ratio between the two measurements of each fiber, the results are consistent within $\pm\SI{0.3}{\percent}$.}
\end{center} 

\subsection{Software}
\label{sec:software}

The software for the test bench is developed under Windows using C++. It can be divided into three hierarchies as follows:
\begin{enumerate}
	\item \textit{Device abstraction}, which not only serves as an interface to the hardware, but also handles the abstraction of different types of devices. 
	\item \textit{Framework libraries}, which defines a general testing procedure and provides utility classes for configuration and management.
	\item \textit{User interface}, which provides command line based or graphical executable for user interaction. 
\end{enumerate}

Instead of developing a dedicated program each time a hardware changes, an abstraction of the devices is adopted to separate the testing procedure from hardware implementation details. 
Abstract classes are defined for the four types of essential equipments as shown in the rounded boxes in Fig.~\ref{fig:FIG1}.
New hardware of each type only needs to inherit from the corresponding abstract class and implement its interface methods and then registered in the singleton device management class \textit{PTDeviceManager}, leaving all other part of the software unchanged.
	
Built upon the abstract device interface, a general testing framework is defined as shown in Fig.~\ref{fig:FIG4}.
\textit{PTVProgram} represents the measurement for a specific characteristic of PMT, such as cathode uniformity, gain and so on.
\textit{PTVTest} is a subunit of \textit{PTVProgram}, which encapsulates the real device operations performed under a specific condition.
A \textit{PTVProgram} may consist of a series of \textit{PTVTest}s, which are invoked sequentially in a test loop.
For example, in cathode uniformity measurement, the stepping motor will move to a series of positions and the PMT response will be recorded by the DAQ at each position.
Here, device operations performed at each position constitute a \textit{PTVTest} and tests at all positions constitute a \textit{PTVProgram}.
Additional operations may be added in the \textit{PreTest} and \textit{PostTest} methods of \textit{PTVProgram}, which will be invoked before and after the test loop respectively.
\textit{PTVProgram}s of various testing objectives are finally chained together to constitute a complete characterization of PMT.

\end{multicols}
\ruleup
\begin{center}
	\includegraphics[width=120mm]{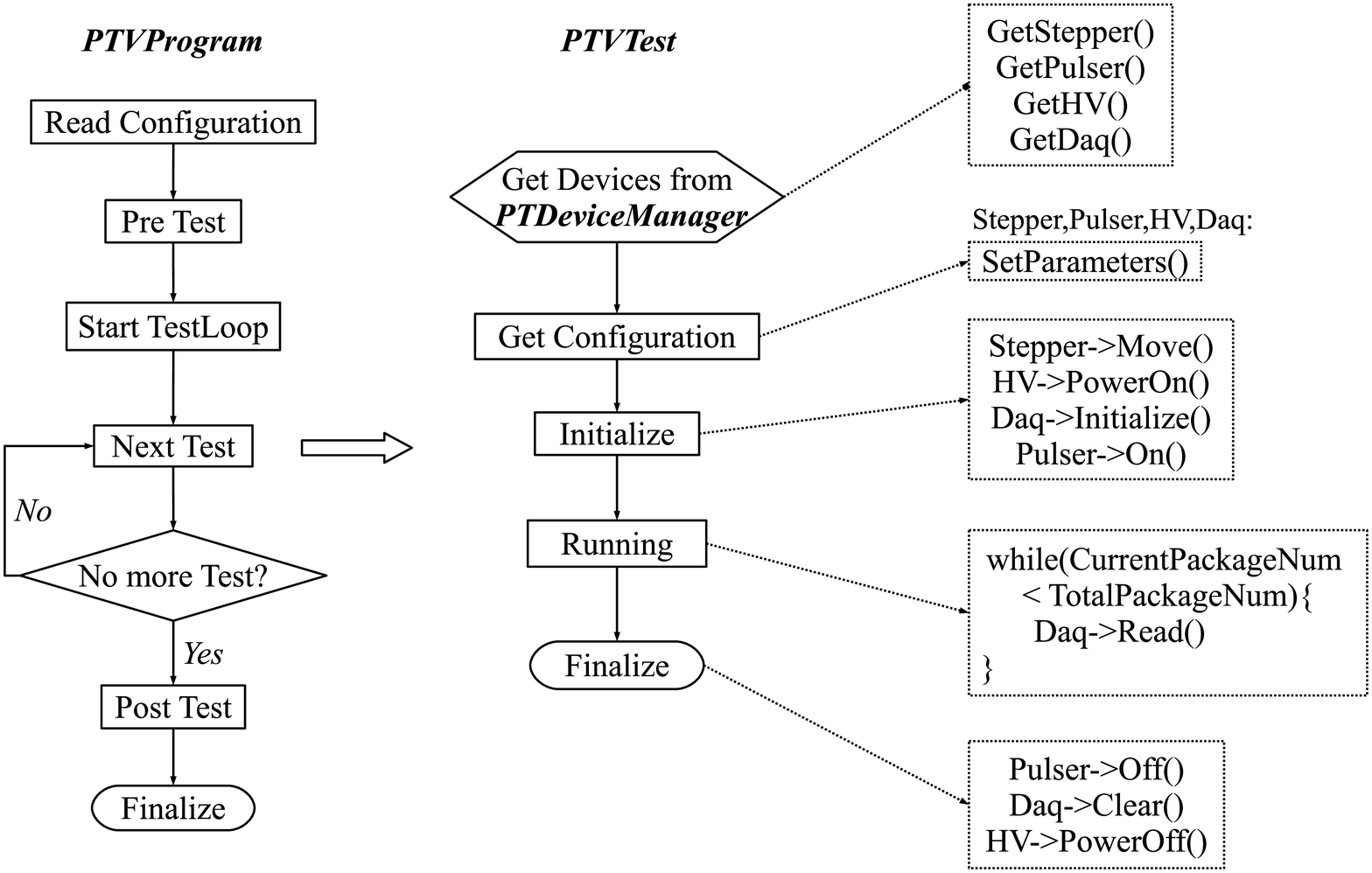}
	\figcaption{\label{fig:FIG4}General PMT testing framework.
		Fake code for a typical \textit{PTVTest} implementation is presented as an example.}
\end{center}
\ruledown
\begin{multicols}{2}
	
For easy usage, a light-weight user interface based on PDCurses~\citep{pdcurses} has also been developed.
This program features in device control, status monitoring and information logging.
A decoding function from binary file to ROOT file~\citep{root} is also incorporated into it for online monitoring.
However, more detailed analysis are considered project-specific and not included in the program. 

\section{Application in the construction of DAMPE-PSD}
\label{sec:application}

PSD is a large-area plastic scintillator array and uses Hamamatsu R4443 PMT for scintillation light detection.
It aims for high energy e/$\gamma$ discrimination as well as charge measurement up to Z=20 by measuring the deposited energy, and no timing information is needed.
To cover the large dynamic range, two dynodes, 5 and 8, are readout for each PMT, 
and the signals are processed by a highly-sensitive ASIC chip (VA160~\citep{va160}, \SI{-3}{\pico\coulomb}$\sim$\SI{13}{\pico\coulomb}) following with an ADC of 14~bits resolution~\citep{fee}. 
Both the gain of dynode8 and the gain between dynode8 and dynode5 of each PMT need to be  measured before the installation.

Concrete \textit{PTVProgram}s for gain, dynode8/dynode5 ratio and cathode uniformity measurements have been implemented.
To accommodate the low input range and obtain more realistic results, the ground-test electronics system of PSD~\citep{fee} are utilized instead of the conventional CAMAC system. 
The ground-test system is mainly a copy of the real ones used in space, and a dedicated \textit{PTVDaq} class based on NI-VISA library~\citep{ni_visa} has been implemented for it.

About 20 tubes are tested in a single run, and it takes normally 5 hours for a complete characterization, including 2 hour's warming time. 
28 test runs are performed in about a month, and totally 570 R4443 tubes have been characterized. 
All the data and analysis results are stored in a MySQL database for easy query, and tubes for installation will be selected based on these data.

The selection procedure is not the subject of this article.
Here, only the major results are presented with a focus on the demonstration of the validity of the test bench. 

\subsection{Relative gain of dynode8}
\label{sec:psd_gain}

PSD requires a \SI{25}{\percent} uniformity in all detection units. 
Therefore, a relative measurement of the gain is adopted by comparing the responses of the PMTs at the same input light intensity.
A light source setting with an intermediate light intensity is selected for this measurement so that the responses of all the tested PMTs are within the linear range.    
To get the real response to the same light intensity, two corrections are applied to the measured data:
\begin{equation}
A_{corr} = \frac{A_{raw}}{k_{runid}\tau_{fiberid}}
\end{equation} 
where $A_{raw}$ is the mean value of the raw ADC spectrum measured at the specified light source setting,
$k_{runid}$ is the light intensity fluctuation of the light source between different test runs,
$\tau_{fiberid}$ is the light output difference between fibers.
$k_{runid}$ can be calculated at each test run using one of the reference PMTs.
$\tau_{runid}$ is a calibration constant which has been measured in Sec.~2.3.
The relative gain $G_{relative}$ can then be obtained by directly dividing the $A_{corr}$ of the tested PMT by that of a standard PMT.   

7 different high voltages, from \SIrange{700}{1000}{\volt} with a \SI{50}{\volt} step, are scanned to obtain the gain variation as a function of supply voltage for each PMT, and the results are fitted using the power law function~\citep{hamamatsu}.
Based on the fitting result, the relative gain at any supply voltage can be calculated.
Distribution of the relative gain at 850V is shown in Fig.~\ref{fig:FIG5}, where the $A_{corr}$ of all tubes are divided by that of the tube with the smallest gain. 
A maximum of about 5.5 times difference in the gain has been observed.

\begin{center}
	\includegraphics[width=\linewidth]{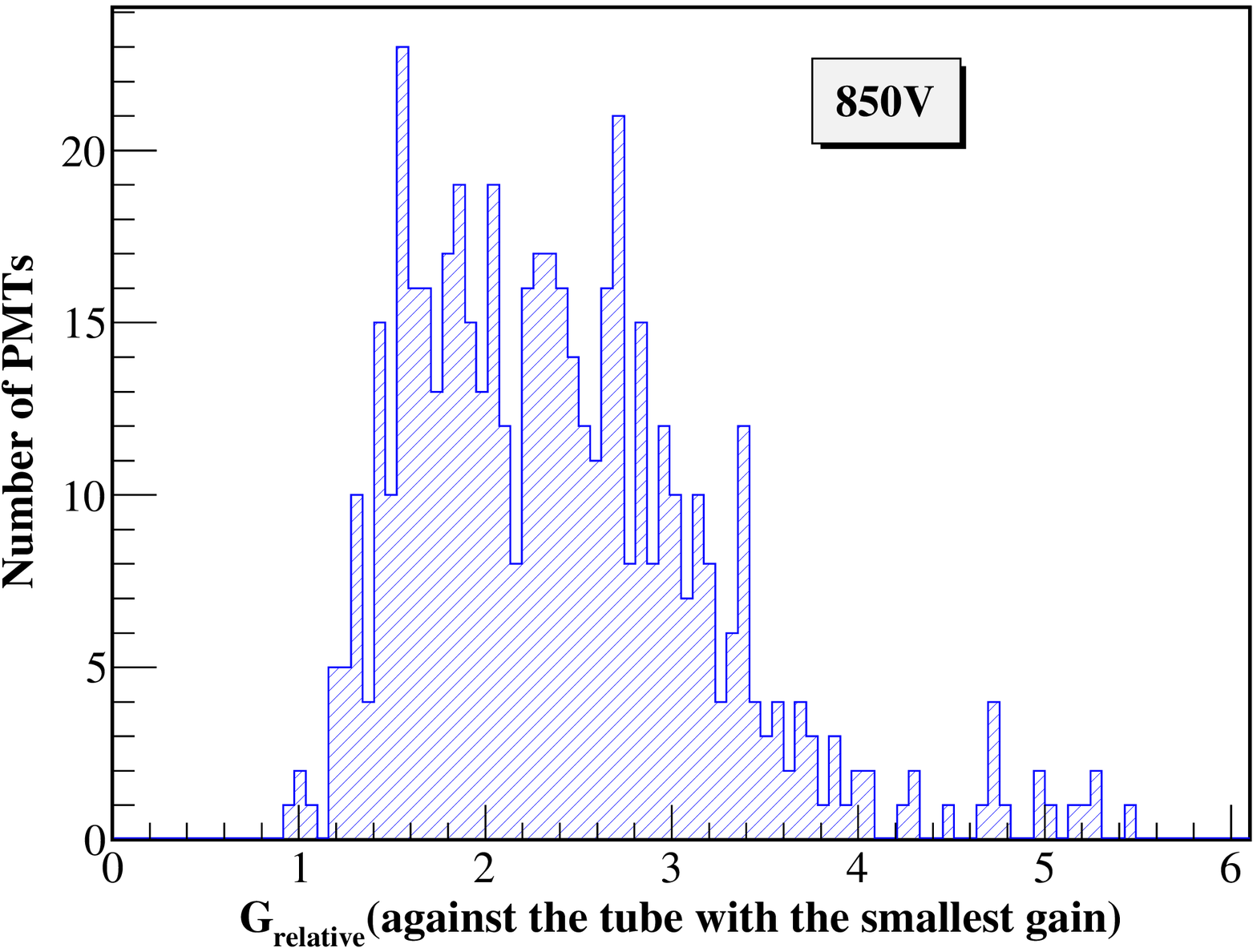}
	\figcaption{\label{fig:FIG5}Relative gain distribution at 850V (normalized to the tube of the smallest gain).}
\end{center}

\subsection{Gain ratio between dynode8 and dynode5}
\label{sec:psd_dy58}

The gain ratio between dynode8 and dynode5 is measured by varying the light intensity in a large range until saturation of the dynode8 signal is observed.
The same procedure is repeated at 7 different supply voltages, from \SIrange{700}{1000}{\volt} with a \SI{50}{\volt} step, to obtain the dynode8/dynode5 dependency on voltage.
As with the gain of dynode8, the dependency can be fitted accurately with a power law function, as shown in the inset graph of Fig.~\ref{fig:FIG6} (a).
The ratio between the gain of dynode8 and dynode5 can then be calculated at any voltage value based on the fitting result.
As an example, the distribution of dynode8/dynode5 ratios at a certain supply voltage is shown in Fig.~\ref{fig:FIG6} (b).
It can be seen that the variance is much smaller than that of gain.

\begin{center}
	\includegraphics[width=\linewidth]{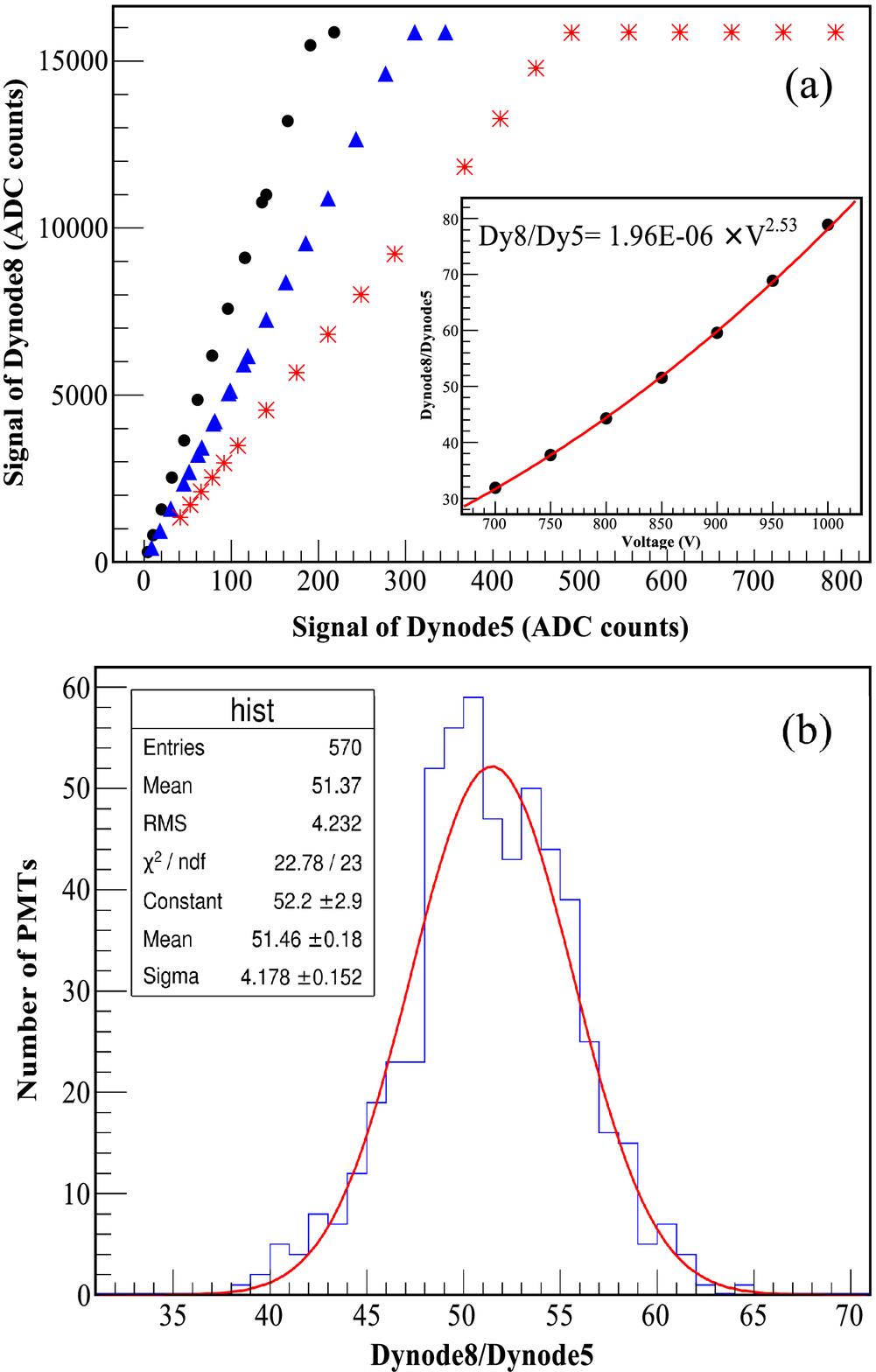}
	\figcaption{\label{fig:FIG6}a) Example of the measurement of dynode8/dynode5 ratio.
		Correlation between dynode5 and dynode8 at 1000V, 850V and 700V is presented, the saturation of dynode8 signal is clearly seen.
		Power law fit to the measured dynode8/dynode5 of this tube at 7 voltage steps is shown in the inset graph. b) Distribution of the dynode8/dynode5 ratio at \SI{820}{\volt} calculated based on the fitted power law function.}
\end{center} 

\subsection{Cathode uniformity}
\label{sec:psd_cathodescan}

The cathode uniformity of R4443 is also checked using the test bench, which has a minimum effective area of about \SI{10}{\milli\meter} in diameter.
The measurement is performed by scanning the input window of R4443 in two perpendicular directions with a step of \SI{1}{\milli\meter}.
At each position, the relative gain is measured at a fixed light source setting according to the method described in Sec.~3.1, and a typical result is shown in Fig.~\ref{fig:FIG7}.

\begin{center}
	\includegraphics[width=\linewidth]{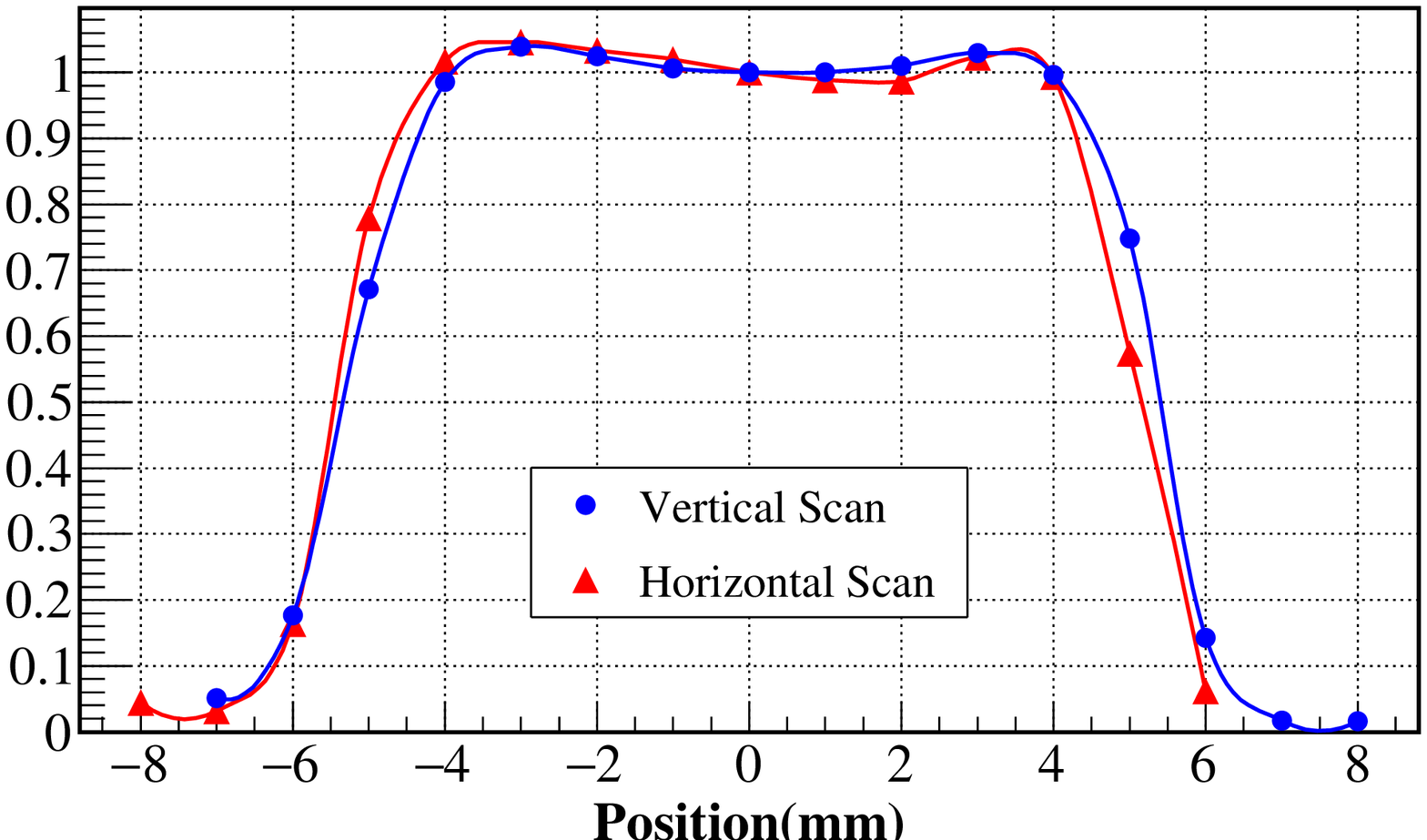}
	\figcaption{\label{fig:FIG7}A typical cathode uniformity of R4443.
		Relative gain at each position is normalized to the center of the input window.}
\end{center} 

The relative gain at each position is proportional to the total efficiency of light transmission, photoelectric conversion and electron collection at this point.
A rapid drop of this efficiency is observed at the edge of the cathode surface. 
Defining the uniform region as a region with the efficiency fluctuation less than \SI{10}{\percent}, it is found that only \SI{75}{\percent} of the tested PMTs can have a uniform region larger than \SI{9}{\milli\meter} in diameter.

\subsection{Stability of the test bench}
\label{sec:stability}

The stability of the test bench during the PMT characterization for DAMPE-PSD can be extracted using the two fixed reference PMTs.

By checking the response of the reference PMTs with the same light pulser setting at different test runs, a maximum variation of \SI{4}{\percent} in the light intensity of the LED is observed during a period of about one month.
Light intensity fluctuation can be corrected using the method described in Sec.~3.1.
As two reference PMTs exists, this method can be validated by using one of them for correction and then checking the relative gain of the other. 
After the correction, stability of the light source is controlled within $\pm\SI{0.5}{\percent}$ as shown in Fig.~\ref{fig:FIG8}.

\begin{center}
	\includegraphics[width=\linewidth]{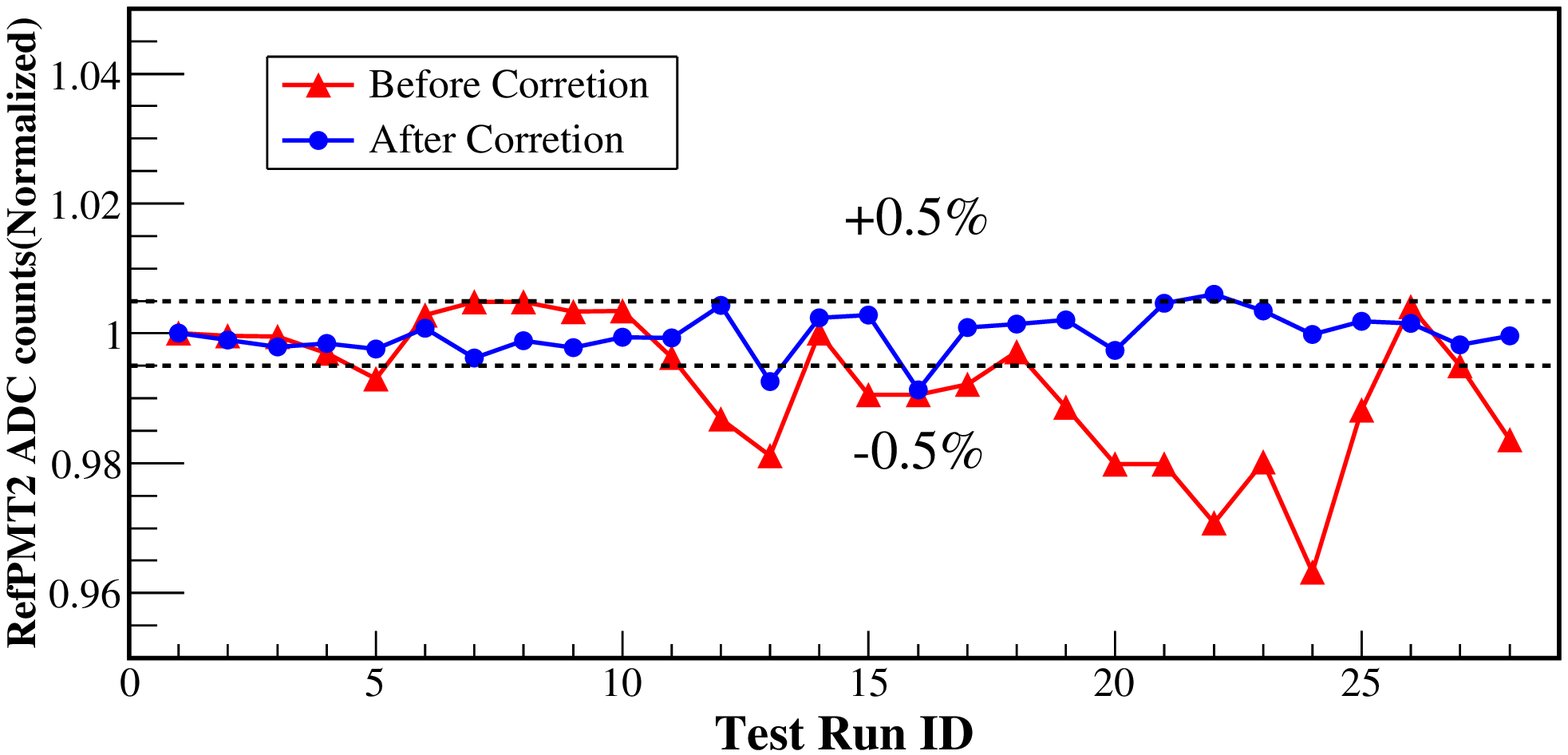}
	\figcaption{\label{fig:FIG8}Stability of the LED monitored by reference PMT2 at 900V (other voltage values give the same results). Red line: $A_{raw}$ before light intensity correction. Blue line: $A_{corr}$ using reference PMT1 for light intensity correction. All the data points are all normalized to the first test run. }
\end{center} 

Reference PMTs underwent the same testing procedures as the tubes under test.
Measurement results of the parameters of the reference PMTs in different test runs can be filled together, and the spread of the distribution is an indication of the uncertainty of the testing method.
In this way, the uncertainty of the dynode8/dynode5 measurement is found to be \SI{1.59}{\percent}, and the uncertainty of relative gain measurement is \SI{0.53}{\percent}. 

\section{Summary}
\label{sec:summary}
A versatile PMT test bench system has been developed at IMP, CAS.
It has been used in the construction of DAMPE PSD already, and the results have proved the high reliability of the system.
Considering the flexible and open platform it possesses, the test bench is useful for any other projects that need massive PMT characterization.

\end{multicols}
\vspace{-1mm}
\centerline{\rule{80mm}{0.1pt}}
\vspace{2mm}

\begin{multicols}{2}

\end{multicols}

\end{CJK*}
\end{document}